\documentclass[acmsmall,screen]{acmart}
\setcopyright{none}
\copyrightyear{2025}
\acmYear{2025}
\acmDOI{}

\usepackage{fancyhdr}
\usepackage{natbib}
\usepackage{mdwtab}
\usepackage{hyperref}
\usepackage[capitalize,nameinlink]{cleveref}
\usepackage{textcomp}
\usepackage{stmaryrd}

\usepackage{tikz}
\usetikzlibrary{matrix}

\usepackage{listings}
\lstdefinestyle{scheme}{
  language=Lisp,
  basicstyle=\ttfamily,
  upquote=true,
  columns=flexible,
  mathescape=true,
  breaklines=true,
  deletekeywords={sum,let,map,cons,null,and,or,cond},
  keepspaces=true,
}
\usepackage{ebproof}
\newcommand{\wmid}{\;\mid\;}
\newcommand{\Rplus}{\mathbb{R}_{\ge0}^\infty}

\title[Committing to the bit: Relational programming with semiring arrays and SAT solving]
      {Committing to the bit: \\Relational programming with semiring arrays and SAT solving}
\author{Dmitri Volkov}
\email{dvolkov@iu.edu}
\author{Yafei Yang}
\email{yafyang@iu.edu}
\author{Chung-chieh Shan}
\email{ccshan@iu.edu}
\affiliation{
  \institution{Indiana University}
  \city{Bloomington}
  \country{USA}}

\begin{document}

\settopmatter{printacmref=false}
\settopmatter{printfolios=true}
\renewcommand\footnotetextcopyrightpermission[1]{}
\pagestyle{fancy}
\fancyfoot{}
\fancyfoot[R]{miniKanren'25}
\fancypagestyle{firstfancy}{
  \fancyhead{}
  \fancyhead[R]{miniKanren'25}
  \fancyfoot{}
}
\makeatletter
\let\@authorsaddresses\@empty
\makeatother

\begin{abstract}
    We propose semiringKanren, a relational programming language where each relation expression denotes a semiring array.
    We formalize a type system that restricts the arrays to finite size.
    We then define a semantics that is parameterized by the semiring that the arrays draw their elements from.
    We compile semiringKanren types to bitstring representations.
    For the Boolean semiring, this compilation enables us to use an SAT solver to run semiringKanren programs efficiently.
    We compare the performance of semiringKanren and faster miniKanren for solving Sudoku puzzles.
    Our experiment shows that semiringKanren can be a more efficient variant of miniKanren.
\end{abstract}

\maketitle
\thispagestyle{firstfancy}

\section{Introduction}

The evaluation of a miniKanren program traditionally searches for solutions \emph{top-down}: invoking a goal that is a conjunction or disjunction causes its subgoals to be invoked in turn.
On one hand, top-down evaluation can efficiently explore possibilities while avoiding useless swaths of the search space, such as when another conjunct already failed or when another disjunct already succeeded.
On the other hand, not only the time and space usage but also the termination behavior of top-down search can depend on the order of subgoals.
Consequently, the illusion that goals denote relations can be hard to maintain in practical programs.

It is useful to generalize relational programming from the Boolean semiring to other semirings.
For example, relations can be generalized to probability distributions and quantum circuits \citep{abo-khamis-convergence-jacm,green-provenance}.
These generalizations motivate alternative semantics for relational programs where solutions are computed \emph{bottom-up} from parts of a relation.

To flesh out the idea of bottom-up evaluation, we propose a toy relational language called \emph{semiringKanren}\footnote{semiringKanren repository: \url{https://github.com/sporkl/semiringkanren}}.
Like miniKanren and relational algebra, semiringKanren is a small language whose expressions are built up using operations such as conjunction, disjunction, and projection (\lstinline|fresh|).
But as its name suggests, semiringKanren computes n-dimensional semiring arrays, which generalize relations over finite domains from the Boolean semiring to other semirings.
Each semiring array element can be thought of as a weight of a domain element.
Because we only consider finite domains, our algebraic data types do not allow recursion.
But we allow recursive \emph{calls}.
They are easy to implement by numerical fixed-point, and our example programs show how they can be useful.

Naive evaluation of semiringKanren programs can be quite inefficient.
We expect ample optimization opportunities for solvers to exploit.
As a first step in this direction, we compile semiringKanren programs over the Boolean semiring into SAT problems that can be fed to off-the-shelf solvers.

\section{Example programs}

We introduce semiringKanren by examples.
A semiringKanren program consists of a \emph{query} supported by zero or more global \emph{relation definitions}.

\subsection{Relations}

A relation takes zero or more arguments, each a typed variable, and assigns a \emph{weight} to the tuple of argument values.
For instance, an unfair coin can be modeled by a relation \lstinline|unfair| that takes one variable \lstinline|coin|, whose type is \lstinline|(Sum Unit Unit)|.
\begin{lstlisting}
(defrel (unfair (coin : (Sum Unit Unit)))
  ...)
\end{lstlisting}
This \lstinline|Sum| type has two values, which we write as \lstinline|(left ())| and \lstinline|(right ())|. 
The type \lstinline|Unit| has just one value~\lstinline|()|.

If we wanted to assign the weight 0.5 to both values, we would simply fill in the \lstinline|defrel| body above with the primitive relation expression \lstinline|(factor 0.5)|, which always assigns the weight 0.5\@.
But to model an unfair coin, we need to distinguish \lstinline|left| from \lstinline|right|, using the primitive relations \lstinline|lefto| and \lstinline|righto| respectively.
The relation expression \lstinline|(lefto coin u)| assigns the weight 1 when \lstinline|coin| is equal to \lstinline|(left u)|, and 0 otherwise.
The variable \lstinline|u| of type \lstinline|Unit| can be bound using \lstinline|fresh|.
So the relation expression
\begin{verbatim}
(fresh ((u : Unit)) (lefto coin u))
\end{verbatim}
assigns the weight 1 when \lstinline|coin| is a \lstinline|left| value (there is only one) and 0 otherwise.

To add and multiply weights, we use the relation combinators \lstinline|disj| and \lstinline|conj|.
Putting it together, we define the \lstinline|unfair| relation as follows:
\begin{lstlisting}
(defrel (unfair (coin : (Sum Unit Unit)))
  (disj (conj (factor 0.7) (fresh ((u : Unit)) (lefto coin u)))
        (conj (factor 0.3) (fresh ((u : Unit)) (righto coin u)))))
\end{lstlisting}
This relation assigns the weight 0.7 to \lstinline|(left ())| and 0.3 to \lstinline|(right ())|.

\subsection{Queries}

A query might postulate two independent flips of an \lstinline{unfair} coin and ask about the probabilities of the cases where the two flips agree.
The following code expresses this query as an implicit conjunction of three relations.
The first two conjuncts invoke \lstinline|unfair| with different arguments.
The third conjunct uses primitive equality~\lstinline|==|, which requires two variables of the same type.
This requirement guarantees that \lstinline|==| returns a relation over finite domains.
\begin{lstlisting}
(run ((coin1 : (Sum Unit Unit))
      (coin2 : (Sum Unit Unit)))
  (unfair coin1)
  (unfair coin2)
  (== coin1 coin2))
\end{lstlisting}
The result is this array of weights:
\begin{center}
    \begin{tabular}{l|cc}
    \hlx{v}
      & \lstinline|(left ())|
      & \lstinline|(right ())|
    \\\hlx{vhv}
        \lstinline|(left ())|
      & 0.49
      & 0\hphantom{.00}
    \\
        \lstinline|(right ())|
      & 0\hphantom{.00}
      & 0.09
    \\\hlx{v}
    \end{tabular}
\end{center}
The equality conjunct \lstinline|(== coin1 coin2)| zeroes out the weights where \lstinline|coin1| and \lstinline|coin2| disagree, so the total weight is less than~1.
To compute the total weight as our query, we can use \lstinline|fresh|, which denotes summation:
\begin{lstlisting}
(run ()
  (fresh ((coin1 : (Sum Unit Unit))
          (coin2 : (Sum Unit Unit)))
    (unfair coin1)
    (unfair coin2)
    (== coin1 coin2)))
\end{lstlisting}
The result is 0.58\@, a rank-0 array (in other words, a scalar).

\subsection{Semirings}

In general, each semiringKanren program can draw weights from a given \emph{commutative semiring}~$\mathbb{K}$.
A~commutative semiring is a set of elements equipped with two associative and commutative operations $+$ and~$\times$, which have identities $0$ and~$1$ respectively and obey distributivity.

So far we have been using the nonnegative reals $\mathbb{R}_{\ge0}$ as our semiring.
We extend $\mathbb{R}_{\ge0}$ to include positive infinity $+\infty$:
\begin{equation}
    \Rplus = \mathbb{R}_{\ge0}\cup\{+\infty\}.
\end{equation}
Then we can take the least fixed points later \citep{droste-semirings}.
Another useful commutative semiring is the Booleans equipped with logical operations $\lor$ (whose identity is~$\bot$) and~$\land$ (whose identity is~$\top$).
We represent this semiring as $\mathbb{B}$.

\subsection{Recursion}
\label{s:recursion}

As illustrated above, our arrays are indexed by the typed values that each variable can range over.
Both sum and product types are supported, but recursive types are not, because we want our arrays to remain finite in size.

On the other hand, recursive calls can work: A relation can be defined by invoking itself.
For instance, a fair coin can be simulated by flipping an unfair coin repeatedly \citep{vonneumannVariousTechniquesUsed1951}:
Flip the unfair coin twice and see if the two flips have the same result.
If they agree, then repeat.
Otherwise, report the first flip as the overall result.
This can be modeled by a recursive relation \lstinline|fair|:
\begin{lstlisting}
(defrel (fair (coin : (Sum Unit Unit)))
  (fresh ((coin1 : (Sum Unit Unit))
          (coin2 : (Sum Unit Unit)))
    (unfair coin1)
    (unfair coin2)
    (disj (conj (== coin1 coin2) (fair coin))
          (conj (=/= coin1 coin2) (== coin coin1)))))
\end{lstlisting}
Here \lstinline{=/=} is the primitive inequality, which requires two variables of the same type.
As with \lstinline{==}, this requirement guarantees that \lstinline|=/=| returns a relation over finite domains.
The array denoted by \lstinline|fair| is the least fixed point
\begin{center}
    \begin{tabular}{cc}
        \lstinline|(left ())| & \lstinline|(right ())|
    \\\hlx{vhv}
        0.5 & 0.5
    \end{tabular}.
\end{center}

Another classic use of recursion in relational programming is to express the transitive closure of a relation.
To demonstrate this, let's encode a graph whose nodes are integers between 0 and~3:
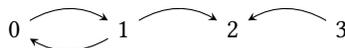
\begin{center}
    \begin{tikzpicture}[>=stealth]
        \node (m) [matrix of nodes, column sep=3em] { 0 & 1 & 2 & 3 \\ };
        \draw [->, bend left] (m-1-1) to (m-1-2);
        \draw [->, bend left] (m-1-2) to (m-1-1);
        \draw [->, bend left] (m-1-2) to (m-1-3);
        \draw [<-, bend left] (m-1-3) to (m-1-4);
    \end{tikzpicture}
\end{center}
The type of an integer between 0 and 3 can be represented as %
\texttt{(Sum Unit (Sum Unit (Sum Unit Unit)))}.
We abbreviate this type as \lstinline|Num|, and abbreviate its four values
\begin{lstlisting}
(left ())
(right (left ()))
(right (right (left ())))
(right (right (right ())))
\end{lstlisting}
as 0, 1, 2, and 3 respectively.
The graph is a binary relation among nodes:
\begin{lstlisting}
(defrel (graph (x : Num) (y : Num))
  (disj (conj (== x $0$) (== y $1$))
        (conj (== x $1$) (== y $0$))
        (conj (== x $1$) (== y $2$))
        (conj (== x $3$) (== y $2$))))
\end{lstlisting}
Reachability can then be expressed as a recursive relation:
\begin{lstlisting}
(defrel (connect (x : Num) (y : Num))
  (disj (graph x y)
        (fresh ((z : Num))
          (conj (connect x z)
                (connect z y)))))
\end{lstlisting}
Querying this recursive relation
\begin{lstlisting}
(run ((x : Num) (y : Num))
  (connect x y))
\end{lstlisting}
produces a different array depending on which semiring is used.
In the Boolean semiring $\mathbb{B}$, we end up testing reachability:
\begin{center}
    \begin{tabular}{l|cccc}
    \hlx{v}
      & 0 & 1 & 2 & 3
    \\\hlx{vhv}
    0 & $\top$ & $\top$ & $\top$ & $\bot$
    \\
    1 & $\top$ & $\top$ & $\top$ & $\bot$
    \\
    2 & $\bot$ & $\bot$ & $\bot$ & $\bot$
    \\
    3 & $\bot$ & $\bot$ & $\top$ & $\bot$
    \\\hlx{v}
    \end{tabular}
\end{center}
In the semiring $\Rplus$, we end up counting paths, and the cycle in the graph gives rise to an infinite number of paths:
\begin{center}
    \begin{tabular}{l|cccc}
    \hlx{v}
      & 0 & 1 & 2 & 3
    \\\hlx{vhv}
    0 & $+\infty$ & $+\infty$ & $+\infty$ & $0$
    \\
    1 & $+\infty$ & $+\infty$ & $+\infty$ & $0$
    \\
    2 & $0$ & $0$ & $0$ & $0$
    \\
    3 & $0$ & $0$ & $1$ & $0$
    \\\hlx{v}
    \end{tabular}
\end{center}
To measure the shortest paths instead of counting the number of paths, we can switch to the \emph{tropical} semiring, where addition is $\min$ (whose identity is $+\infty$) and multiplication is $+$ (whose identity is~$0$).
An element of this semiring can be thought of as the length or cost of an edge or path.
Let each edge in our graph have length~$10$:

\noindent
\begin{minipage}{\linewidth}
\begin{lstlisting}
(defrel (graph (x : Num) (y : Num))
  (disj (conj (factor 10) (== x $0$) (== y $1$))
        (conj (factor 10) (== x $1$) (== y $0$))
        (conj (factor 10) (== x $1$) (== y $2$))
        (conj (factor 10) (== x $3$) (== y $2$))))
\end{lstlisting}
\end{minipage}
Then some shortest paths have length~$10$ whereas others have length~$20$:
\begin{center}
    \begin{tabular}{l|cccc}
    \hlx{v}
      & 0 & 1 & 2 & 3
    \\\hlx{vhv}
    0 & $20$ & $10$ & $20$ & $+\infty$
    \\
    1 & $10$ & $20$ & $10$ & $+\infty$
    \\
    2 & $+\infty$ & $+\infty$ & $+\infty$ & $+\infty$
    \\
    3 & $+\infty$ & $+\infty$ & $10$ & $+\infty$
    \\\hlx{v}
    \end{tabular}
\end{center}

\subsection{Sudoku puzzle}

As a larger example, we show how to encode a $4\times4$ Sudoku puzzle.
The relation \lstinline{valid$_{4}$} takes four arguments of
type \lstinline{Num}. It requires that any two of them have different
values.
It is defined using the primitive inequality relation \lstinline|=/=|.
\begin{lstlisting}
(defrel (valid$_4$ (a : Num) (b : Num) (c : Num) (d : Num))
  (=/= a b) (=/= a c) (=/= a d) (=/= b c) (=/= b d) (=/= c d))
\end{lstlisting}
The relation \lstinline{sudoku$_{4\times{4}}$} takes sixteen arguments of type \lstinline{Num}, which represent the 16 numbers we put on a $4\times{4}$ board.
It requires that there is no duplicate in the four numbers we put into each of the four rows, each of the four columns, and each of the four $2\times{2}$ quadrants.
We invoke \lstinline{valid$_4$} on the corresponding arguments to satisfy these requirements.
\begin{lstlisting}
(defrel (sudoku$_{4\times{4}}$
          (a : Num) (b : Num) (c : Num) (d : Num)
          (e : Num) (f : Num) (g : Num) (h : Num)
          (i : Num) (j : Num) (k : Num) (l : Num)
          (m : Num) (n : Num) (o : Num) (p : Num))
  (valid$_4$ a b c d) (valid$_4$ a b e f) (valid$_4$ a e i m)
  (valid$_4$ e f g h) (valid$_4$ c d g h) (valid$_4$ b f j n)
  (valid$_4$ i j k l) (valid$_4$ i j m n) (valid$_4$ c g k o)
  (valid$_4$ m n o p) (valid$_4$ k l o p) (valid$_4$ d h l p))
\end{lstlisting}
We run the following query to get the solution to a Sudoku puzzle.
\begin{lstlisting}
(run (          (b : Num)           (d : Num)
      (e : Num) (f : Num) (g : Num) (h : Num)
      (i : Num) (j : Num) (k : Num) (l : Num)
      (m : Num)           (o : Num)          )
  (sudoku$_{4\times{4}}$ $3$ b $0$ d e f g h i j k l m $0$ o $2$))
\end{lstlisting}
The result is a $4^{12}$ array that is all zero except for the weight~$1$ at
$e=l=0$, $d=f=k=m=1$, $b=g=i=2$, and $h=j=o=3$.

\section{Syntax and typing}

\begin{figure}
\begin{array}{@{}Tlr@{\;}c@{\;}l@{}}
    Programs & P &  ::=  & \mathit{Rel} \ldots \; \mathit{Query}
\\  Relations& \mathit{Rel} & ::= & \texttt{(defrel (}R\texttt{ (}x\texttt{ :\ }\tau\texttt{)}\:\ldots\texttt{) }g\texttt{) }
\\  Queries  & \mathit{Query} & ::= & \texttt{(run ((}x\texttt{ :\ }\tau\texttt{)}\:\ldots\texttt{) }g\texttt{)}
\\  Goals    & g &  ::=  & \texttt{(conj }g\texttt{ }g\texttt{)}
\\           &   & \wmid & \texttt{(disj }g\texttt{ }g\texttt{)}
\\           &   & \wmid & \texttt{(factor }r\texttt{)}
\\           &   & \wmid & \texttt{(fresh ((}x\texttt{ :\ }\tau\texttt{)) }g\texttt{)}
\\           &   & \wmid & \texttt{(}R\texttt{ }x\:\ldots\texttt{)}
\\           &   & \wmid & \texttt{(== }x\texttt{ }y\texttt{)}
\\           &   & \wmid & \texttt{(=/= }x\texttt{ }y\texttt{)}
\\           &   & \wmid & \texttt{(soleo }x\texttt{)}
\\           &   & \wmid & \texttt{(lefto }x\texttt{ }y\texttt{)}
\\           &   & \wmid & \texttt{(righto }x\texttt{ }y\texttt{)}
\\           &   & \wmid & \texttt{(pairo }x\texttt{ }y\texttt{ }z\texttt{)}
\\  Types & \tau &  ::=  & \texttt{Unit}
\\           &   & \wmid & \texttt{(Sum }\tau\texttt{ }\tau\texttt{)}
\\           &   & \wmid & \texttt{(Prod }\tau\texttt{ }\tau\texttt{)}
\\  Relation names & R
\\  Variables & x,y,z
\\  Weights  & r & \in & \mathbb{K}
\end{array}
\caption{Syntax}
\label{fig:syntax}
\end{figure}

We give the syntax of semiringKanren in \cref{fig:syntax}.
For simplicity, our formalization omits some syntactic sugar:
\begin{itemize}
\item \lstinline{run} takes just one goal.
\item \lstinline{conj} and \lstinline{disj} take exactly two goals.
\item \lstinline{fresh} binds one variable at a time.
\item Arguments must be variables, not literal values or value constructors. For instance, the goal \lstinline{(== x $0$)} in \cref{s:recursion} above desugars to
\begin{lstlisting}
(fresh ((z : Num))
  (conj (fresh ((u : Unit))
          (conj (soleo u)
                (lefto z u)))
        (== x z)))
\end{lstlisting}
\end{itemize}

\begin{figure}
\begin{centering}
\lineskiplimit 2ex
\lineskip 2ex
\emergencystretch 0pt
\everymath{\xspaceskip0pt \spaceskip0pt \normalbaselines}
\xspaceskip 3em plus.5em minus1.25em
\spaceskip 2em plus.5em minus1.25em
\linepenalty 1000
\hbadness 10000
\begin{prooftree}
  \hypo{\Gamma; \Delta \vdash g_1 \quad \Gamma; \Delta \vdash g_2}
  \infer1{\Gamma; \Delta \vdash \texttt{(conj }g_1\texttt{ }g_2\texttt{)}}
\end{prooftree}
\begin{prooftree}
  \hypo{\Gamma; \Delta \vdash g_1 \quad \Gamma; \Delta \vdash g_2}
  \infer1{\Gamma; \Delta \vdash \texttt{(disj }g_1\texttt{ }g_2\texttt{)}}
\end{prooftree}
\begin{prooftree}
  \hypo{r \in \mathbb{K}}
  \infer1{\Gamma; \Delta \vdash \texttt{(factor }r\texttt{)}}
\end{prooftree}
\begin{prooftree}
  \hypo{\Gamma;\!\;\Delta,\!\;x:\tau \vdash g}
  \infer1{\Gamma; \Delta \vdash \texttt{(fresh ((}x\texttt{ :\ }\tau\texttt{)) }g\texttt{)}}
\end{prooftree}
\begin{prooftree}
  \hypo{(R:\tau_1,\dotsc,\tau_n\rightarrow) \in \Gamma}
  \hypo{x_1:\tau_1 \in \Delta \enspace \ldots \enspace x_n:\tau_n \in \Delta}
  \infer2{\Gamma; \Delta \vdash \texttt{(}R\texttt{ }x_1\ldots x_n\texttt{)}}
\end{prooftree}
\begin{prooftree}
  \hypo{x:\tau \in \Delta }
  \hypo{y:\tau \in \Delta}
  \infer2{\Gamma; \Delta \vdash \texttt{(== }x\texttt{ }y\texttt{)}}
\end{prooftree}~%
\begin{prooftree}
  \hypo{x:\tau \in \Delta}
  \hypo{y:\tau \in \Delta}
  \infer2{\Gamma; \Delta \vdash \texttt{(=/= }x\texttt{ }y\texttt{)}}
\end{prooftree}
\begin{prooftree}
  \hypo{x:\texttt{Unit} \in \Delta}
  \infer1{\Gamma; \Delta \vdash \texttt{(soleo }x\texttt{)}}
\end{prooftree}~%
\begin{prooftree}
  \hypo{x:\texttt{(Sum }\tau_1\texttt{ }\tau_2\texttt{)} \in \Delta}
  \hypo{y:\tau_1 \in \Delta}
  \infer2{\Gamma; \Delta \vdash \texttt{(lefto }x\texttt{ }y\texttt{)}}
\end{prooftree}~%
\begin{prooftree}
  \hypo{x:\texttt{(Sum }\tau_1\texttt{ }\tau_2\texttt{)} \in \Delta}
  \hypo{y:\tau_2 \in \Delta}
  \infer2{\Gamma; \Delta \vdash \texttt{(righto }x\texttt{ }y\texttt{)}}
\end{prooftree}
\begin{prooftree}
  \hypo{x:\texttt{(Prod }\tau_1\texttt{ }\tau_2\texttt{)} \in \Delta}
  \hypo{y:\tau_1 \in \Delta}
  \hypo{z:\tau_2 \in \Delta}
  \infer3{\Gamma; \Delta \vdash \texttt{(pairo }x\texttt{ }y\texttt{ }z\texttt{)}}
\end{prooftree}

\end{centering}
\caption{Type system}
\label{fig:type-system}
\end{figure}

The type system is shown in \cref{fig:type-system}.
The typing judgment $\Gamma;\Delta\vdash g$ means that $g$ is a well-typed goal,
where $\Gamma$ gives the types of relations and $\Delta$ gives the types of variables.
Each entry in $\Delta$ has the form $x:\tau$, where $x$ is a variable
(bound by \lstinline{run} or \lstinline{fresh}) and $\tau$ is its type.
Each entry in $\Gamma$ has the form $(R:\tau_1,\dotsc,\tau_n\rightarrow)$,
which means that $R$ is an $n$-ary relation whose arguments have types
$\tau_1,\dotsc,\tau_n$.

Each relation definition
\begin{equation}
\label{e:rel}
\mathit{Rel} = \texttt{(defrel (}R\texttt{ (}x_1\texttt{ :\ }\tau_1\texttt{)}\ldots\texttt{(}x_n\texttt{ :\ }\tau_n\texttt{)}\texttt{) }g\texttt{) }
\end{equation}
purports a relation type $(R:\tau_1,\dotsc,\tau_n\rightarrow)$.
To type-check a complete program
\begin{equation}
    P = \mathit{Rel}_1 \ldots \mathit{Rel}_m \; \mathit{Query},
\end{equation}
we first let $\Gamma$ be the $m$ relation types purported by each of $\mathit{Rel}_1 \ldots \mathit{Rel}_m$.
Then for each $i=1,\dotsc,m$, we check the relation definition $\mathit{Rel}_i$ of form~\eqref{e:rel} by~checking
\begin{equation}
    \Gamma;\; x_1:\tau_1,\dotsc,x_n:\tau_n \vdash g.
\end{equation}
Finally, we also check the query
\begin{equation}
\label{e:query}
    \mathit{Query} = \texttt{(run ((}x_1\texttt{ :\ }\tau_1\texttt{)}\ldots\texttt{(}x_n\texttt{ :\ }\tau_n\texttt{)) }g\texttt{)}
\end{equation}
by~checking
\begin{equation}
    \Gamma;\; x_1:\tau_1,\dotsc,x_n:\tau_n \vdash g.
\end{equation}

\section{Semantics}
\label{s:semantics}

Each type $\tau$ denotes a finite set of values $\llbracket\tau\rrbracket$:
\begin{equation}
\label{e:values}
\begin{split}
    \llbracket \texttt{Unit} \rrbracket &= \{\texttt{()}\}
\\
    \llbracket \texttt{(Sum }\tau_1\texttt{ }\tau_2\texttt{)} \rrbracket &= \{\,\texttt{(left }v\texttt{)}\mid v\in\llbracket\tau_1\rrbracket\,\} \cup \{\,\texttt{(right }v\texttt{)}\mid v\in\llbracket\tau_2\rrbracket\,\}
\\
\llbracket \texttt{(Prod }\tau_1\texttt{ }\tau_2\texttt{)} \rrbracket &= \{\,\texttt{(}v_1\texttt{ .\ }v_2\texttt{)}\mid v_1\in\llbracket\tau_1\rrbracket,\!\;v_2\in\llbracket\tau_2\rrbracket\,\}
\end{split}
\end{equation}
The values in each set can be ordered lexicographically, so we can use values as array indices.

\begin{figure}
\newcommand\weightOf[1]{\llbracket #1 \rrbracket (\eta;\delta)}
\newcommand\indicator[1]{\bigl\{
    \begin{smallmatrix}
        1 & \text{if }#1 \hfill \\
        0 & \text{otherwise} \hfill
    \end{smallmatrix}}
\begin{align*}
    \weightOf{ \texttt{(conj }g_1\texttt{ }g_2\texttt{)} }
    & = \weightOf{ g_1 } \times \weightOf{ g_2 }
\\
    \weightOf{ \texttt{(disj }g_1\texttt{ }g_2\texttt{)} }
    & = \weightOf{ g_1 } + \weightOf{ g_2 }
\\
    \weightOf{ \texttt{(factor }r\texttt{)} }
    & = r
\\
    \weightOf{ \texttt{(fresh ((}x\texttt{ :\ }\tau\texttt{)) }g\texttt{)} }
    & = \textstyle\smash[b]{ \sum_{v\in\llbracket\tau\rrbracket} \llbracket g \rrbracket (\eta; \delta\{x\mapsto v\}) }
\\
    \weightOf{ \texttt{(}R\texttt{ }x_1\ldots x_n\texttt{)} }
    & = \eta(R)\bigl(\delta(x_1),\dotsc,\delta(x_n)\bigr)
\\
    \weightOf{ \texttt{(== }x\texttt{ }y\texttt{)} }
    & = \indicator{\delta(x) = \delta(y)}
\\
    \weightOf{ \texttt{(=/= }x\texttt{ }y\texttt{)} }
    & = \indicator{\delta(x) \ne \delta(y)}
\\
    \weightOf{ \texttt{(soleo }x\texttt{)} }
    & = \indicator{\delta(x) = \texttt{()}}
\\
    \weightOf{ \texttt{(lefto }x\texttt{ }y\texttt{)} }
    & = \indicator{\delta(x) = \texttt{(left }\delta(y)\texttt{)}}
\\
    \weightOf{ \texttt{(righto }x\texttt{ }y\texttt{)} }
    & = \indicator{\delta(x) = \texttt{(right }\delta(y)\texttt{)}}
\\
    \weightOf{ \texttt{(pairo }x\texttt{ }y\texttt{ }z\texttt{)} }
    & = \indicator{\delta(x) = \texttt{(}\delta(y)\texttt{ .\ }\delta(z)\texttt{)}}
\end{align*}
\caption{Semantics}
\label{fig:semantics}
\end{figure}

To define the denotational semantics of a semiringKanren program, \cref{fig:semantics} interprets each derivation of a typing judgment $\Gamma;\Delta\vdash g$ as a function that takes a relation environment~$\eta$ (that matches the relation type environment~$\Gamma$) and a value environment~$\delta$ (that matches the value type environment~$\Delta$) and returns a weight (that belongs to the semiring~$\mathbb{K}$).
In order for $\delta$ to match~$\Delta$, for each $x:\tau\in\Delta$, the value environment~$\delta$ must map the variable~$x$ to an element of~$\tau$.
In order for $\eta$ to match~$\Gamma$, for each $(R:\tau_1,\dotsc,\tau_n\rightarrow)$, the relation environment~$\eta$ must map the relation name~$R$ to a rank-$n$ array whose elements are indexed by $\llbracket\tau_1\rrbracket\times\dotsb\times\llbracket\tau_n\rrbracket$ and drawn from~$\mathbb{K}$.
In short, the meaning of a goal maps the relations it uses to an array whose indices range over the values of the variables it uses.

Each line of \cref{fig:semantics} can be understood as defining an array indexed by~$\delta$ and parameterized over~$\eta$.
For instance, \lstinline{conj} and \lstinline{disj} denote elementwise multiplication and addition, and \lstinline{(factor $r$)} produces an array filled with the constant~$r\in\mathbb{K}$.
More interestingly, \lstinline{fresh} sums a rank-$(n+1)$ array along one dimension to produce a rank-$n$ array.
Equality \lstinline{(== $x$ $y$)} denotes a diagonal array (unless $x$ and~$y$ are the same variable, in which case it denotes the array filled with the multiplicative identity~$1$).

We define the denotation of a complete program by taking the least fixed point of a monotone function~$F$ from relation environments to relation environments.
To~define this monotone function~$F$, we let
\begin{equation}
    F(\eta)(R)(v_1,\dotsc,v_n) = \llbracket g \rrbracket(\eta; \{x_1\mapsto v_1,\dotsc,x_n\mapsto v_n\})
\end{equation}
for each relation name~$R$ defined by \eqref{e:rel} in the program.
Now let $\eta$ be the least fixed point of~$F$, and suppose the query in the program is~\eqref{e:query}.
Then the program simply denotes the rank-$n$ array that assigns to each index $(v_1,\dotsc,v_n)$ the weight
\begin{equation}
    \llbracket g \rrbracket(\eta; \{x_1\mapsto v_1,\dotsc,x_n\mapsto v_n\}).
\end{equation}

\section{Compilation}
\label{s:compilation}

The denotational semantics just described can also be regarded as a naive way to compute the result of a program.
To~validate our work, we have implemented the semantics as executable code.
However, such a naive implementation is wholly unsuitable for practical use.
As a more efficient alternative when the semiring is the Boolean semiring $\mathbb{B}$, we compile semiringKanren programs to SAT problems that can then be fed to an off-the-shelf SAT solver.

The compiler from semiringKanren to SAT works in three stages.
Only the third stage is specific to the Boolean semiring, so this work might help us compile probabilistic or quantum programs in the future.

\subsection{Unrolling calls}

The first stage is to unroll every relation call $\texttt{(}R\texttt{ }x\:\ldots\texttt{)}$ by inlining the definition of the relation $R$.
To force this unrolling to terminate, we impose a limit on the call depth.
If this limit is hit (which will happen if the source program is recursive), then we replace the call by failure.
Consequently, the result computed by the SAT solver might be less than the correct result (that is, some trues might become falses).

\subsection{Bitstring representation}

The second stage of the compiler commutes with the first.
It produces intermediate semiringKanren code in which the only sum type is the type of a binary bit:
\begin{equation}
    \tau^* ::= \texttt{Unit}
    \wmid \texttt{(Sum Unit Unit)}
    \wmid \texttt{(Prod }\tau^*\texttt{ }\tau^*\texttt{)}
\end{equation}
A value of such a type can be thought of as a bitstring, so we call it a \emph{bitstring value} of a \emph{bitstring type}. 
The key to this stage is to replace each type~$\tau$ by a bitstring type~$\tau^*$, as follows:
\begin{equation}
\label{e:compile_adt}
\begin{split}
\texttt{Unit}^*
& = \texttt{Unit}
\\
\texttt{(Sum }\tau_1\texttt{ }\tau_2\texttt{)}^*
& = \begin{cases}
        \texttt{(Sum Unit Unit)}
        & \text{if }\tau_1=\tau_2=\texttt{Unit}
    \\
        \texttt{(Prod (Sum Unit Unit) }\tau_1^*\texttt{)}
        & \text{if }|\llbracket\tau_1^*\rrbracket| > |\llbracket\tau_2^*\rrbracket|
    \\
        \texttt{(Prod (Sum Unit Unit) }\tau_2^*\texttt{)}
        & \text{otherwise}
    \end{cases}
\\
\texttt{(Prod }\tau_1\texttt{ }\tau_2\texttt{)}^*
& = \texttt{(Prod }\tau_1^*\texttt{ }\tau_2^*\texttt{)}
\end{split}
\end{equation}
The essence of this type compilation is to encode a sum type by a tagged union---a bitstring whose first bit is a tag.
The rest of the bitstring is just long enough to accommodate either variant.
For instance, the type
\begin{equation}
\label{e:three}
    \texttt{(Sum Unit (Sum Unit Unit))}
\end{equation}
represents an integer between 0 and 2.
When compiled, it becomes the bitstring type
\begin{equation}
\label{e:two-bits}
    \texttt{(Prod (Sum Unit Unit) (Sum Unit Unit))}
\end{equation}
that represents 2 bits.

Note that the original type \eqref{e:three} has 3 values, whereas the bitstring type \eqref{e:two-bits} has 4 values, according to the semantics \eqref{e:values}.
This growth is to be expected because the number of values in any bitstring type can only be a power of~2.
Yet we want to preserve the meaning of expressions such as
\begin{lstlisting}
(fresh ((x : (Sum Unit (Sum Unit Unit))))
  (factor 1))
\end{lstlisting}
(which should mean 3\@, not 4) and
\begin{lstlisting}
(fresh ((a : (Sum Unit (Sum Unit Unit)))
        (b : (Sum Unit (Sum Unit Unit)))
        (c : (Sum Unit (Sum Unit Unit)))
        (d : (Sum Unit (Sum Unit Unit))))
  (=/= a b) (=/= a c) (=/= a d) (=/= b c) (=/= b d) (=/= c d))
\end{lstlisting}
(which should mean 0\@, or false).
To this end, the compiler inserts a conjunction immediately under each variable binding (\texttt{fresh} or \texttt{run}) to enforce that the bitstring represents a valid value.
For example, in the case of \eqref{e:three}, the conjunction rules out 1~of the 4 possible values of the bitstring type~\eqref{e:two-bits}. 

{\sloppy
Separately, when compiling \lstinline{(lefto $x$ $y$)} where the sum type of~$x$ is represented by \texttt{(Prod~(Sum~Unit~Unit)~$\tau_{2}^{*}$)} in~\eqref{e:compile_adt}, or dually when compiling \texttt{(righto }$x$\texttt{ }$y$\texttt{)} where the sum type of~$x$ is represented by \texttt{(Prod (Sum Unit Unit) $\tau_1^*$)} in~\eqref{e:compile_adt}, the compiler needs to coerce the type of~$y$ between $\tau_1^*$ and~$\tau_2^*$, so that the produced code stays well-typed.

}

\subsection{Propositional formula}

Once a semiringKanren program uses only bitstring types, it~is easy to convert it to a propositional formula that is satisfiable if and only if the result of the program is true.
This final compiler stage simply recurs over the program syntax. 
The interesting cases below are \texttt{fresh}, \texttt{lefto}, and \texttt{righto}:
\begin{itemize}
    \item \texttt{conj} compiles to~$\land$, and \texttt{disj} compiles to~$\lor$.
    \item \texttt{(factor $r$)} compiles to $\bot$ if $r$ is the additive identity~$0$, or~$\top$ otherwise.
    \item \texttt{fresh} introduces $\log_2 |\llbracket\tau^*\rrbracket|$ propositional variables, then proceeds to compile the body.
    \item \texttt{==}, \texttt{=/=}, and \texttt{pairo} compile to equality ($\leftrightarrow$) between propositional variables.
    \item \texttt{soleo} compiles to~$\top$.
    \item \texttt{lefto} and \texttt{righto} compile simply to testing the first bit of the scrutinized bitstring.
\end{itemize}

\section{Benchmark}

We evaluate the performance of semiringKanren by running Sudoku programs.
We collect Sudoku puzzles of different difficulty levels \citep{SolveWebSudoku2025,4x4SudokuKids2025} and try to solve them in three ways:
\begin{itemize}
    \item Naive implementation of semiring-generic semiringKanren semantics (\cref{s:semantics}) using array operations from the scientific computing package Owl \citep{wangOCamlScientificComputing2025}
    \item Compilation to a propositional formula (\cref{s:compilation}) fed to an off-the-shelf SAT solver \citep{buryMSATOCamlSAT2025}
    \item Faster miniKanren \citep{ballantyneFasterMinikanren2025}
\end{itemize}
We implemented the semiringKanren semantics and compiler in OCaml 4.14.1\@\footnote{semiringKanren benchmark repository: \url{https://github.com/chihyang/tensorkanren_benchmark}}.
We performed all experiments on a Ubuntu 22.04 machine with 1.6 GHz CPU and 20 GB of RAM\@.

\begin{table}
\caption{Time (in seconds) to solve Sudoku puzzles using semiringKanren and faster miniKanren}
\label{table:perf}
\centering
\begin{tabular}{lccc}
  \toprule
  Sudoku program & semiringKanren semantics & SAT solver & Faster miniKanren \\
  \midrule
  $4 \times 4$         & Out of memory & \hphantom{0}0.227  & 0.746     \\
  $9 \times 9$, medium & Out of memory & 25.814 & Timeout   \\
  $9 \times 9$, hard   & Out of memory & 28.043 & Timeout   \\
  $9 \times 9$, expert & Out of memory & 72.088 & Timeout   \\
  \bottomrule
\end{tabular}
\end{table}

The timing results are shown in \cref{table:perf}.
Only the SAT solver solves all the puzzles in reasonable time.
The semiringKanren semantics implementation runs out of memory for all test programs.
Faster miniKanren solves the $4 \times 4$ Sudoku puzzle in reasonable time, but fails to solve all the other puzzles within 2 minutes.

\begin{table}
\caption{Problem sizes when Sudoku programs are compiled to SAT}
\label{table:sat-size}
\centering
\begin{tabular}{lccc}
  \toprule
  Sudoku program & \# SAT variables & \# SAT clauses & Average SAT clause size \\
  \midrule
  $4 \times 4$         & \hphantom{00}2278 & \hphantom{00}14360 & 2.377 \\
  $9 \times 9$, medium & 290208  & 2301808   & 2.484 \\
  $9 \times 9$, hard   & 293512  & 2317791   & 2.484 \\
  $9 \times 9$, expert & 291782  & 2309539   & 2.484 \\
  \midrule
  $9 \times 9$, hand encoding \citep{lynceSudokuSATProblem2006} & \hphantom{000}810 & \hphantom{000}8829 & 2.189 \\
  \bottomrule
\end{tabular}
\end{table}

Although the SAT solver outperforms the other implementations, there is a lot of room for further improvement, given that the compiled SAT problems are far from the most compact encoding \citep{lynceSudokuSATProblem2006}. This can be seen in the compiled SAT problem sizes in \cref{table:sat-size}.
The bottom row shows the SAT encoding designed specifically for Sudoku by \citet{lynceSudokuSATProblem2006}.
Our compiler produces SAT encodings that are more than two orders of magnitude bigger, in terms of the number of variables and clauses.

\section{Related work}

Our language is typed, whereas original miniKanren is untyped.
There have been many attempts to embed miniKanren into a typed programming language:
OCanren in OCaml \citep{kosarevTypedEmbeddingRelational2018},
canrun\_rs in Rust \citep{simmlerTgechoCanrun_rs2025},
gominikanren in Go \citep{schulzeAwalterschulzeGominikanren2025},
klogic in Kotlin \citep{kamenevKlogicMiniKanrenKotlin2023a},
and typedKanren in Haskell \citep{kudasovTypedKanrenStaticallyTyped2024}.
These implementations need to type the values passed to a relation.
But they still follow a top-down evaluation strategy as the original miniKanren does \citep{friedmanReasonedSchemer2018}.

There are a variety of ways in which an SAT or SMT solver could be used to enhance relational programming.
One way is to use an SAT solver to prune the search in miniKanren \citep{ballantyneMichaelballantyneMkcdcl2024}.
More generally, SMT solvers have been used to optimize queries in other relational programming languages such as Datalog \citep{wangOptimizingRecursiveQueries2022} and SQL \citep{chuCosetteAutomatedProver2017}.
Alternatively, a relational programming language can expose an interface to an underlying SMT solver so that programs can invoke the solver and state constraints directly \citep{zuckerMakingMiniKanrenUsing2021,chenChansey97ClprosetteminiKanrenCLPRosette}.
In contrast, semiringKanren does not expose such an interface to the programmer.

\section{Future work}


Our implementation currently works on an AST representation of semiringKanren programs, which could use a friendlier frontend. It would be interesting to try to make the embedding more shallow.

Our SAT solver implementation currently only returns one solution. It would be more useful to produce multiple solutions.

It remains to extend our semantics and implementation to infinite domains, including recursive data types, while preserving bottom-up evaluation.
It also remains to extend our semantics and implementation to polymorphic relations.
Either extension would make it easier to express applications such as arithmetic.

It is not obvious whether semiringKanren is Turing-complete or not; certain semirings may enable or disable this property.


\bibliographystyle{plainnat}
\bibliography{main}

\end{document}